\documentclass[aps,onecolumn,preprint,superscriptaddress,longbibliography]{revtex4-1}

\usepackage[dvips]{hyperref,color,graphicx}      %dvips or pdftex
\usepackage{amsmath,amssymb,amstext}
\usepackage{nicefrac}

\hypersetup{
  naturalnames=true,
  colorlinks=true,
  linkcolor=blue,
  pdfpagemode=UseNone,
  pdfstartview=FitH,
  pdftitle={Collective strong coupling of cold potassium atoms in a ring cavity},
  pdfauthor={J. Goldwin},
  pdfsubject={Collective strong coupling of cold potassium atoms in a ring cavity},
  pdfkeywords={Cold atoms, Cavity QED},
}

\begin{document}

\newcommand{\ket}[1]{| #1 \rangle}
\renewcommand{\Re}{\mbox{Re}}
\renewcommand{\Im}{\mbox{Im}}
\newcommand{\nwsearrow}{\mathrel{\text{$\nwarrow$\llap{$\searrow$}}}}
\newcommand{\xxx}{\textcolor{red}{XXX}}

\def\UoB{
	School of Physics and Astronomy, 
	University of Birmingham, 
	Edgbaston, Birmingham B15 2TT, 
	United Kingdom
}
\def\USP{
	Instituto de F\'{i}sica de S\~{a}o Carlos, 
	Universidade de S\~{a}o Paulo, 13560-970 
	S\~{a}o Carlos, SP, Brazil
}

%\preprint{}

\title{Collective strong coupling of cold potassium atoms\\in a ring cavity}

\author{R. Culver}
\affiliation{\UoB}
\author{A. Lampis}
\affiliation{\UoB}
\author{B. Megyeri}
\affiliation{\UoB}
\author{K. Pahwa}
\affiliation{\UoB}
\author{L. Mudarikwa}
\altaffiliation{Present address: TSC Inspection Systems, Davy Avenue, Knowlhill, Milton Keynes MK5 8PB, United Kingdom.}
\affiliation{\UoB}
\author{M. Holynski}
\affiliation{\UoB}
\affiliation{\USP}
\author{Ph.~W. Courteille}
\affiliation{\USP}
\author{J. Goldwin}
\altaffiliation{Corresponding author.}
\affiliation{\UoB}

%\thanks{}
\date{\today}

\begin{abstract}
We present experiments on ensemble cavity quantum electrodynamics with cold potassium atoms in a high-finesse ring cavity. Potassium-39 atoms are cooled in a two-dimensional magneto-optical trap and transferred to a three-dimensional trap which intersects the cavity mode. The apparatus is described in detail and the first observations of strong coupling with potassium atoms are presented. Collective strong coupling of atoms and light is demonstrated via the splitting of the cavity transmission spectrum and the avoided crossing of the normal modes.
\end{abstract}

%\pacs{42.50.Pq, 42.50.Lc, 07.77.Gx}
\maketitle

\section{Introduction}

The interactions between a single photon and atom in free space are typically very weak. Jaynes and Cummings showed that the coupling matrix element, which we denote $\hbar g$, depends inversely on the square root of the volume occupied by the electromagnetic field \cite{JaynesCummings}. Therefore it is advantageous for studies of cavity quantum electrodynamics (CQED) to confine the atom and light within an optical microcavity \cite{YeCavs,CavsClassical}. For initial conditions with the atom in its excited state and a photon number state $\ket{n}$ of the cavity field, the Rabi oscillation frequency is equal to $2g(1+n)^{1/2}$. For a small enough cavity, even the vacuum ($n=0$) Rabi oscillation frequency can exceed the atomic and photonic decoherence rates ($\gamma$ and $\kappa$, respectively in this work), and oscillatory excitation exchange between the atom and light can occur. The condition that $g$ is large enough that vacuum Rabi oscillations persist over several cycles before damping is conventionally taken as the definition of the strong coupling regime of CQED. 

The presence of vacuum Rabi oscillations can be detected through the spectral splitting of a weakly probed system \cite{Eberly83}. The experimental observation of normal-mode splitting of cavity transmission spectra with a single or a few atoms was an important milestone in the historical development of CQED \cite{Raizen89,Mossberg90,TRK92}. More recently a wide range of experiments have begun to study CQED with large atom number. The multi-atom extension of the Jaynes-Cummings Hamiltonian was provided by Tavis and Cummings \cite{TavisCummings}, and later extended by Agarwal \cite{Agarwal84} to include damping. Ensemble CQED differs from single-atom CQED in some important ways. From a practical viewpoint, the vacuum Rabi frequency increases with atom number according to $g\to g\,N^{1/2}$, relaxing the technical constraints on the optical cavity design. More fundamentally, a wealth of new physics can arise if the atomic density distribution extends in space across several optical wavelengths. This is associated with effective long-range interactions between atoms mediated by the quantum optical field \cite{Courteille06,RitschRMP}. Collective vacuum Rabi splitting in particular has been central to studies of optomechanical effects in ring cavities \cite{Hemmerich06}, atomic spin squeezing \cite{Kasevich06,Thompson11}, cavity linewidth control \cite{Zhu10}, CQED with multiple atomic states \cite{Barrett11} and cavity modes \cite{Renzoni13}, and cavity Rydberg polaritons \cite{Simon16}.

Here we present the first demonstration of collective strong coupling of cold potassium atoms, using a high-finesse ring cavity. Compared with more commonly used elements such as rubidium or caesium, potassium offers a choice of stable bosonic and fermionic isotopes with varying and tuneable atom-atom interactions \cite{FeshbachRMP}. The relatively small hyperfine splittings also make it potentially easier to reach a regime where multiple atomic states are mixed in the presence of strong light-matter coupling \cite{Birnbaum06,Barrett11}. The outline of this paper is as follows. In Section~\ref{sec:apparatus} we describe our experimental apparatus, including the vacuum system (\ref{sec:vacuum}), the laser system (\ref{sec:lasers}), and the ring cavity itself (\ref{sec:cavity}). In Section~\ref{sec:science} we demonstrate strong coupling on the D$_1$ lines of potassium-39, first through the observation of the vacuum Rabi splitting and its dependence on atom number, and second through the avoided crossing of the normal mode resonances across a range of cavity and probe laser detunings. A collective cooperativity of $C=g^2N/(\kappa\gamma)>100$ is achieved, implying a large effective susceptibility for future studies of nonlinear optics with large dispersion.

\section{Apparatus\label{sec:apparatus}}

\subsection{Vacuum system\label{sec:vacuum}}

\begin{figure*}[ht]
\centering
\includegraphics[height=4cm]{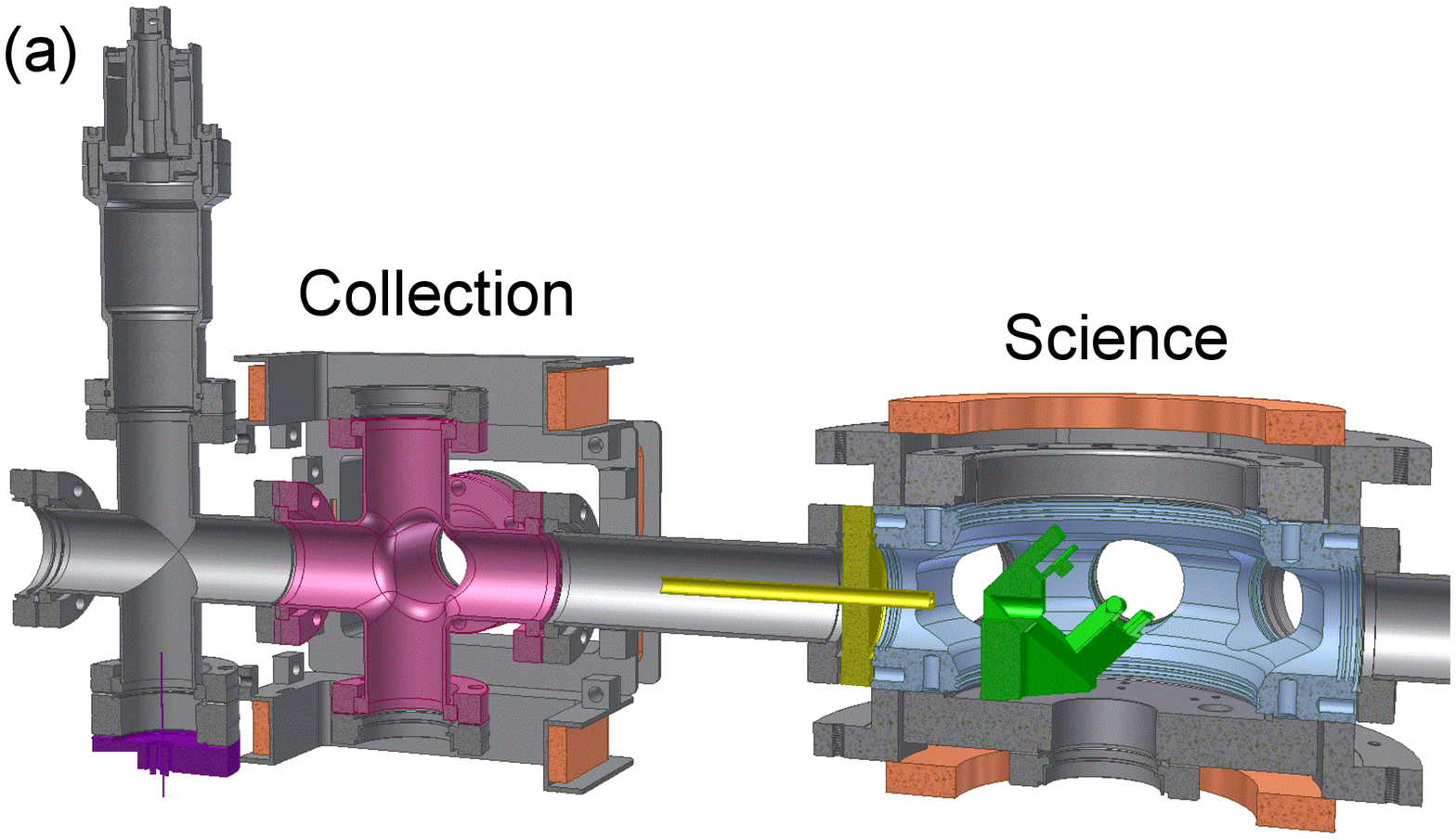}\hfill
\includegraphics[height=4cm]{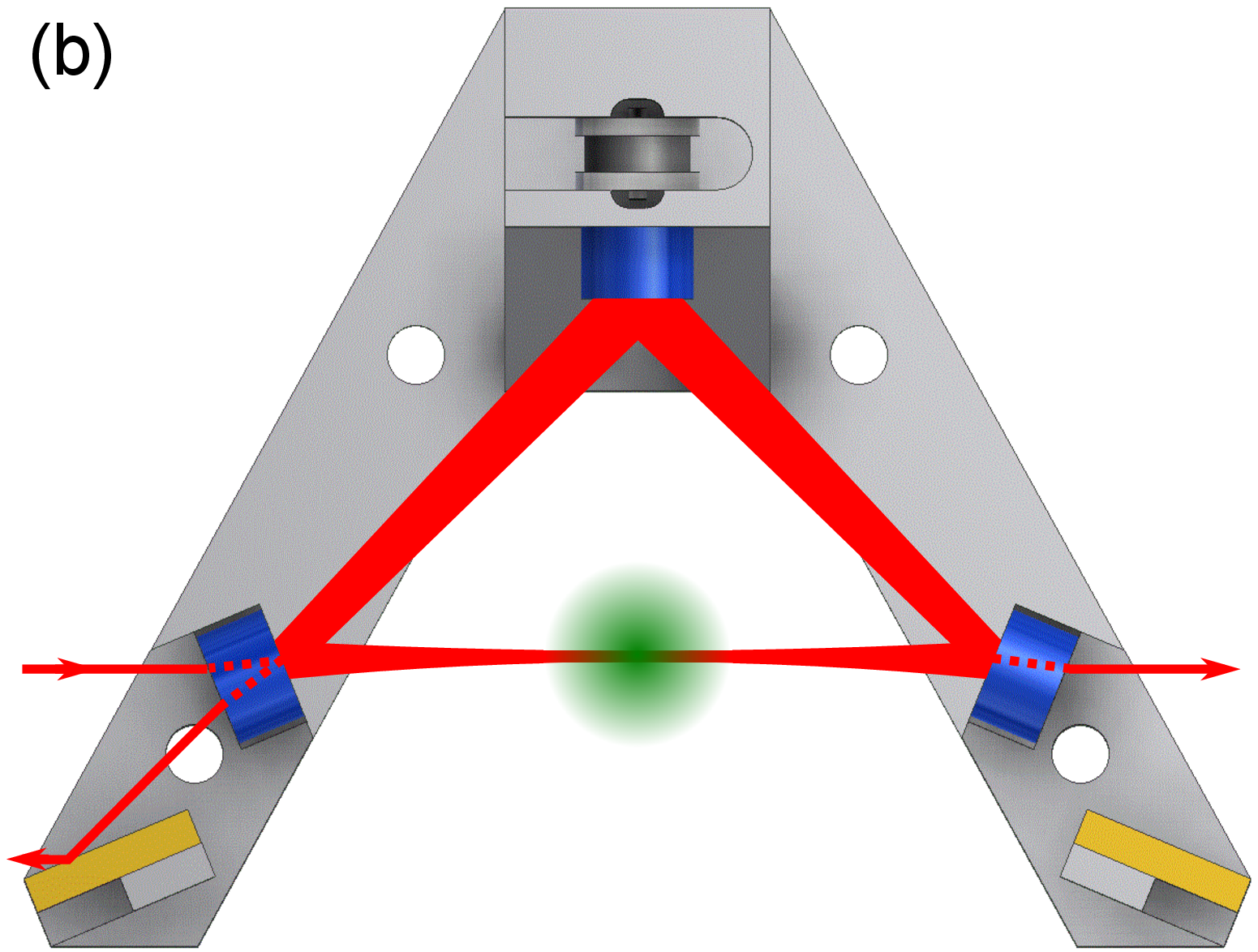}\hfill
\includegraphics[height=4cm]{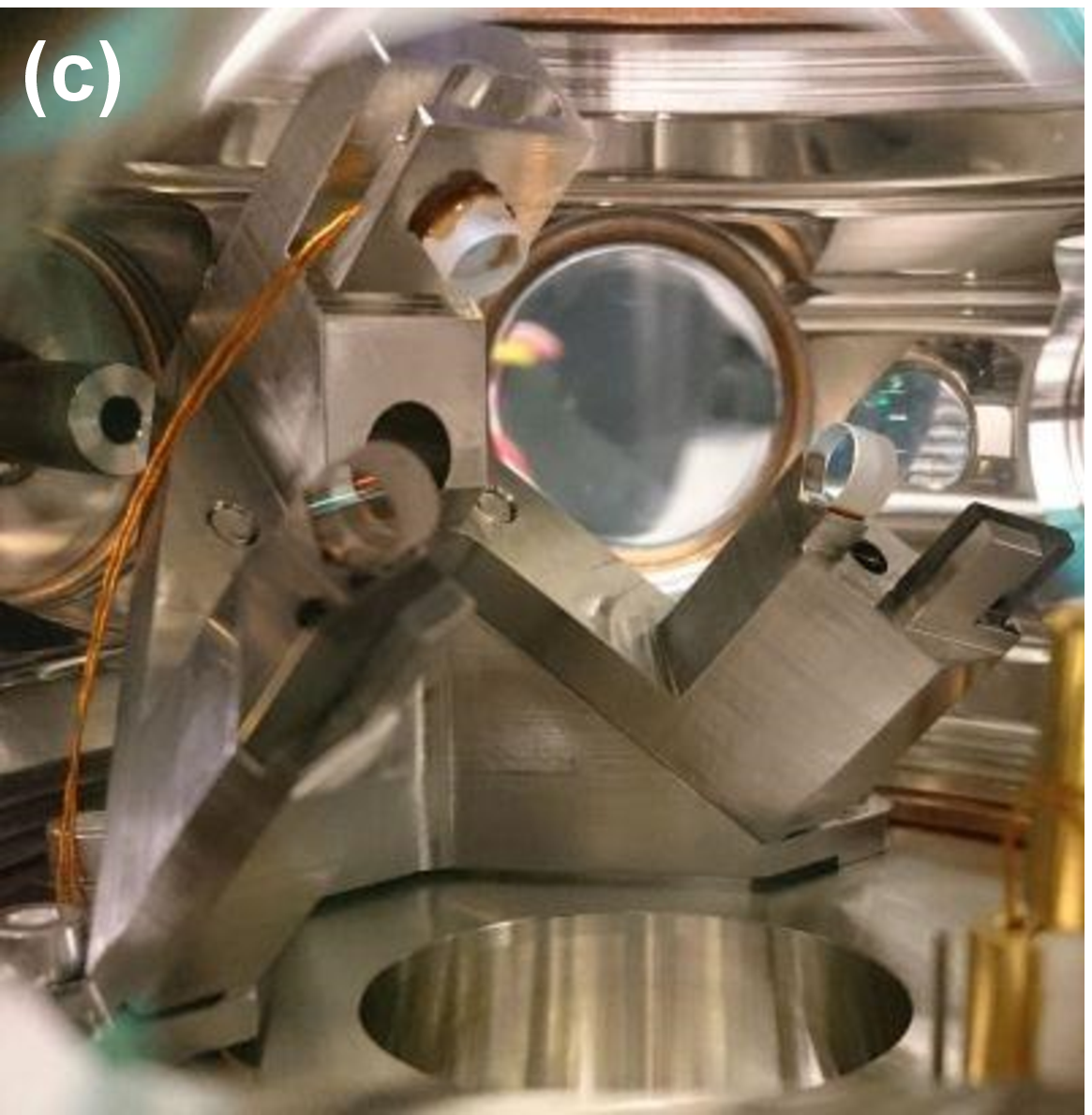}\hfill
\caption{Experimental apparatus. (a) Vacuum system, with collection chamber on the left (pink) and science chamber on the right (light blue). The cavity frame and riser are shown in green. For clarity we have omitted an all-metal valve and 20~L/s ion pump after the science chamber. The graphite transfer tube (yellow) is mounted within the nipple joining the two chambers. The dispenser feedthrough is shown in purple, and the 2D- and 3D-MOT coils in copper. (b) Schematic of the ring cavity, as viewed from above the cavity plane. The cavity mirrors are coloured blue and the gold-coated steering mirrors are in yellow; the piezo and ceramic spacers are visible within the flexure. For clarity only the counter-clockwise cavity mode is illustrated (the mode shapes are the same, but the input/output directions are mirrored). The cavity mode and potassium MOT (green) are not to scale. (c) Photograph of the ring cavity in the science chamber, with the cavity plane at $45^\circ$ from horizontal. The end of the transfer tube is visible on the left; atoms emerge from the tube and pass through a hole in the cavity frame on the way to the 3D-MOT. The window in the background and the hole in the bottom flange are for MOT beams.}\label{fig1}
\end{figure*}

An overview of the vacuum system is shown in Fig.~\ref{fig1}(a). The system is split into two main sections: a relatively high-pressure \textit{collection} chamber housing the potassium vapour source, and a low-pressure \textit{science} chamber containing the high-finesse ring cavity. A narrow graphite \textit{transfer tube} (Goodfellow, 494-159-79 \cite{Products}) supports the differential pressure required to keep the science chamber clean. This tube is 100~mm long, with an inner diameter of 3~mm, and is mounted in a stainless steel tube which is welded into a blank ConFlat flange. The transfer tube maintains a calculated pressure ratio of 340:1 between the collection and science chambers. In a first generation apparatus, in which a single chamber housed both the potassium source and the cavity \cite{ThesisPahwa,ThesisMudarikwa}, we achieved collective strong coupling but found that the cavity finesse degraded over the time scale of a few weeks. We have been operating the two-chamber apparatus for around $1\,\nicefrac{1}{2}$ years with no detectable decrease in finesse. 

Potassium atoms are released into the collection chamber from alkali metal dispensers (SAES, K/NF/4.5/25/FT10) mounted on an electrical feedthrough, and aimed at the walls of the surrounding stainless steel cross. The cross is kept heated, along with the rest of the source side of the apparatus, in order to maintain a high enough potassium vapour pressure. Potassium-39 atoms from the thermal background are cooled in a two-dimensional magneto-optical trap (2D-MOT) formed in a standard six-way cross. All windows used in the experiment are Kodial viewports with broadband anti-reflection coatings. A single 20~L/s ion pump is attached to the far end of the science chamber, described below. All-metal valves on both ends allow roughing during bake-down. Atoms from the 2D-MOT are pushed through the transfer tube with a near-resonant laser beam and collected in a 3D-MOT in the science chamber. 

The science chamber comprises a commercial spherical octagon (Kimball Physics, MCF600-SphOct-FC28). A large reducing flange on the bottom holds the cavity frame and a window for passing the vertical 3D-MOT beams. The flange has been modified to provide mounting holes for the ring cavity and to accommodate a welded-in electrical feedthrough for the cavity tuning piezo. The cavity frame was rigidly mounted to the flange in order to reduce long-term drifts to the alignment. However we observe that the stabilized cavity is disturbed by the fast ($\sim100~\mu$s) shut-off of the MOT coils. Although vibration isolation of the coils reduced this effect, it has not been eliminated. We believe that eddy currents induced in the chamber and/or cavity frame are responsible for the remaining disturbance, and would recommend either internally isolating the cavity from the chamber or replacing the steel chamber with a glass one. The 3D-MOT coils themselves are wound from 60 turns of Kapton-insulated copper ribbon wire (High Precision Foils, HP04-252) and attached to the optical bench with Sorbothane vibration isolation; a current of 10~A provides a quadrupole gradient of 8~G/cm in the strong (vertical) direction.

\subsection{Laser system\label{sec:lasers}}

The trapping and cooling laser subsystem employs three home-built external cavity diode lasers of the kind described in \cite{Mudarikwa12}. One laser serves as a master, locked to the potassium-39 hyperfine ground state crossover resonance using sub-Doppler magnetically-induced dichroism \cite{Pahwa12}. A slave laser is offset-locked to the master using a side-of-filter technique \cite{Ritt04}. The master-slave beat note is mixed with a voltage-controlled oscillator (VCO) whose output frequency is tuned with an analogue output from a computer control card (National Instruments, PCI-6733). The slave laser is stabilized near the D$_2$ $F=2\leftrightarrow F'=3$ cooling transition (here $F$ is the total electronic plus nuclear angular momentum, and primes denote excited states), and a fraction of the light is shifted by $2\times227$~MHz with a double-passed acousto-optic modulator (AOM) for repumping on the $F=1\leftrightarrow F'=2$ transition \cite{Potassium}. The cooling and repumping beams are then re-combined and injected into a home-built tapered amplifier (M2K, TA-0765-100043), producing a total output of $\sim 500$~mW. The cooling:repumping power balance is approximately 1:1~before the amplifier and 3:2~after. Some of this light is sent to a second amplifier (New focus TA-7613-P) to provide light for the 2D-MOT. This amplifier was manufactured for 780~nm, but provides $\sim3\times$ gain at 767~nm, which is enough for our experiments. After a fibre 50/50 beam splitter and beam expansion optics, we have two one-inch beams ($1/e^2$ diameter) with peak intensities of $\sim 40~\mathrm{mW/cm}^2$, which are retroreflected for the 2D-MOT. The rest of the light from the first amplifier is shifted up and back down in frequency through a pair of AOMs, providing tuning and fast extinction of the 3D-MOT beams. In this work the two MOTs operate with the same detunings. After fibre coupling and beam expansion we obtain three one-inch diameter beams of $\sim 10~\mathrm{mW/cm}^2$, which are retroreflected for the 3D-MOT. Finally, a second slave laser is offset-locked to the first slave as above to provide a pushing beam for transferring atoms from the 2D-MOT to the 3D-MOT. The pushing beam power is $\sim1$~mW and the $1/e^2$ intensity radius is $\sim 1$~mm. The beam is detuned 18~MHz below the D$_2$ $F=1\leftrightarrow F'=2$~transition and linearly polarized, and is blocked with a shutter during measurements. After blocking the pushing beam, the $1/e$ lifetime of the 3D-MOT exceeds 2~s, which is much longer than the typical experimental cycle time (less than 1~ms).

\begin{figure}[ht]
\centering
\includegraphics[width=0.7\textwidth]{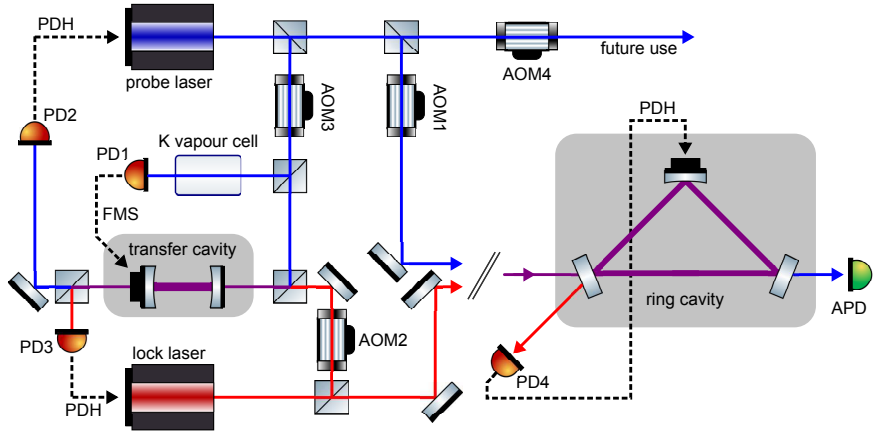}
\caption{Simplified schematic of the cavity lock and probe laser systems. The lock laser operates at 852~nm and the probe laser at 770~nm. AOM: acousto-optic modulator; PD: photodiode; APD: avalanche photodiode; FMS: frequency modulation spectroscopy; PDH: Pound-Drever-Hall. All AOMs are in double-pass configuration. AOM1 tunes the probe beam, whose transmitted power is detected at the APD, AOM2 tunes the ring cavity, and AOM3/AOM4 are reserved for future experiments.}\label{fig2}
\end{figure}

The cavity stabilization and probe laser subsystems, shown schematically in Fig.~\ref{fig2}, use two commercial lasers (Toptica DLPro). One operates around 852~nm, far away from any potassium resonances, and is used for stabilizing the ring cavity length with minimal disturbance to the atoms. The other laser is set throughout this work to probe the potassium D$_1$ transitions at 770~nm, but is capable of reaching the D$_2$ transitions as well. The two wavelengths are selectively combined, split, or blocked using dichroic filters (Thorlabs, FEL0800 and FES0800) and wavelength-specific waveplates (Lens-Optics, W2M15, half-wave at 767~nm and full-wave at 852~nm) and standard polarizing beam splitters.

In order to stabilize both the science cavity and probe laser to arbitrary detunings, we have built a Fabry-Perot \textit{transfer cavity} based on the design in \cite{Budker00}. The design exploits the degeneracy of transverse modes to sub-divide the free spectral range (FSR) into an integer number $r$ of resonances, which are equally spaced by $\mathrm{FSR}/r$. In our case the cavity is $18.8$~cm long, with $r=24$, giving resonances every $33.2$~MHz; the linewidth is $3.8(2)$~MHz at 770~nm and $4.4(2)$~MHz at 852~nm (both half-width at half-maximum). The lasers are current-modulated at $16.6$~MHz to produce sidebands for Pound-Drever-Hall stabilization. The use of a modulation frequency equal to half the mode spacing results in a distinctive square-wave shape of the error signals, with locking points of alternating slopes separated by the modulation frequency \cite{ThesisMudarikwa}. This separation sets the coarse resolution of the laser system. Fine tuning is provided by AOMs which can span neighbouring lock points. Some of the 770~nm light is used to stabilize the transfer cavity itself using sub-Doppler frequency modulation spectroscopy \cite{Mudarikwa12} to control a piezo ring actuator behind one mirror. A small fraction of the 770~nm light is shifted to the D$_1$ $F=1\leftrightarrow F'=2$ transition for probing the cavity, with the rest of the light shifted to $F=2\leftrightarrow F'=2$ for future experiments. We use $250~\mu$W of 852~nm light to stabilize the ring cavity using Pound-Drever-Hall locking, but several mW are available if we wish to produce an intracavity optical dipole trap in the future.

\subsection{Ring cavity\label{sec:cavity}}

In CQED experiments with single atoms in the strong coupling regime, the Fabry-Perot geometry is preferred for geometrical reasons --- it is relatively straightforward to produce a small open mode volume, and therefore large coupling strength $g$, in the gap between a pair of parallel mirrors. In contrast, ensemble CQED relaxes the constraints on mode volume, making ring geometries viable alternatives. The demonstration of collective atomic recoil lasing with cold atoms \cite{CARL} relied intrinsically on the presence of distinct counter-propagating travelling wave modes in a triangular ring cavity. The cavity-enhanced quantum memory of Ref.~\cite{Pan12} also exploited such modes for phase-matched four-wave mixing. Bow-tie cavities have been used for making quantum non-demolition measurements \cite{Bouyer11} and for creating cavity Rydberg polaritons \cite{Simon16}.

Our ring cavity is shown in detail in Fig.~\ref{fig1}(b) and (c). Three mirrors of diameter $6.35$~mm are arranged in a symmetric right-angle triangle with a hypotenuse of nominal length 40~mm in a plane tilted $45^\circ$ from horizontal. The central mirror has a 100~mm radius of curvature (ROC) and is glued with low-outgassing epoxy (Epotek, H74) directly onto the face of a flexure hinge machined into the stainless steel frame. The flexure is driven with a vacuum-compatible piezo actuator (Noliac, NAC2121-H6-C02) which is sandwiched between thin ceramic pieces to electrically insulate the electrodes from the frame. The planar corner mirrors are glued into vee-grooves machined into the frame. They are used for input and output coupling, in conjunction with gold-coated mirrors mounted at right angles to the cavity mirrors to bring counter-propagating pairs of input and output beams parallel. 

The cavity mirrors were sputter coated in a single batch (Layertec, C213A051). The multilayer dielectric coating was designed to produce a finesse of $\sim 1800$ for s-polarized light over the wavelength range 767--852~nm, taking into account the different mirror reflectivities at $45^\circ$ and $22.5^\circ$ angles of incidence. The target power reflectivity at 770~nm was $99.96\%$ for the central mirror and $99.85\%$ for the input-output coupling mirrors, with scattering and absorption losses specified by the manufacturer to be below 100~ppm. When $86.7\%$ of the incident power is matched to a single cavity mode, we observe $41.1\%$ transmission (meaning transmitted power divided by total incident power), and a minimum reflected power of $21.7\%$. The transmitted light is filtered to remove the 852~nm light and the remaining probe light is coupled into a single mode fibre with $\sim 70\%$ efficiency. The light after the fibre is detected with an analog avalanche photodiode (APD) with specified responsivity $>40$~MV/W, noise equivalent power $<7.5~\mathrm{fW}/\sqrt{\mathrm{Hz}}$, and 3-dB bandwidth of 3~MHz (Laser Components UK, LCSA500-03). The empty cavity transmission spectrum has a linewidth of $\sim2\pi\times940$~kHz for s-polarization, and cavity ring-down measurements yield $\kappa=2\pi\times 920(30)$~kHz. The lower reflectivity of the mirrors for p-polarization results in a $\sim5.3\times$ larger linewidth. We restrict ourselves to the higher finesse polarisation for the rest of this work.

\begin{figure}[ht]
\centering
\includegraphics[width=0.5\textwidth]{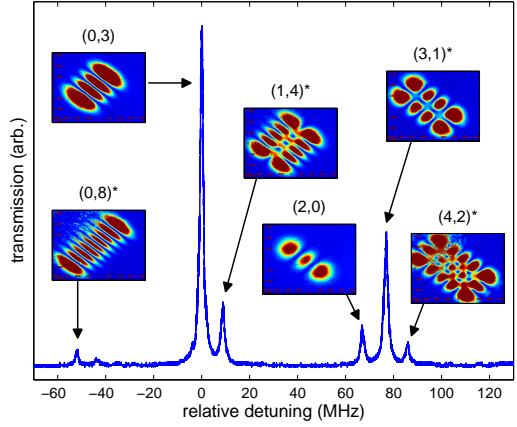}
\caption{Transverse cavity modes. The solid curve shows a number of cavity transmission resonances (without atoms), and insets show the corresponding spatial profiles as imaged onto a camera. The $45^\circ$ tilt of the patterns reflects the orientation of the cavity plane ($\nwsearrow$). The modes are labelled according to transverse indices $(m,n)$, with asterisks denoting a longitudinal index $q$ which is one less than that of the $(0,3)$ mode which is taken as reference.}\label{fig3}
\end{figure}

To characterize the cavity further we exploit the inherent astigmatism of the ring geometry. Because of the $45^\circ$ angle of incidence on the curved mirror, the effective radius of curvature is $R_\parallel = \mathrm{ROC}/\sqrt{2}$ along the tangential plane and $R_\perp=\mathrm{ROC}\sqrt{2}$ in the sagittal plane. This in turn leads to different Gouy phases, splitting the degeneracy of higher-order (transverse) Hermite-Gaussian cavity modes \cite{Lange98}. In Figure~\ref{fig3} we show a transmission spectrum where the incident probe beam has been misaligned deliberately in order to excite numerous transverse modes. For our geometry the resonance frequencies are given by,
\begin{widetext}
\begin{eqnarray}\label{eq:modes}
\omega^q_{m,n} &=& \mathrm{FSR}\left\{q + (m+1/2)\,\frac{\cos^{-1}(1-L/R_\parallel)}{2\pi} + (n+1/2)\,\frac{\cos^{-1}(1-L/R_\perp)}{2\pi} + \frac{1}{4}\,\left[1-(-1)^m\right]\right\}\quad.
\end{eqnarray}
\end{widetext}
Here $(q,m,n)$ are the longitudinal, tangential, and sagittal mode numbers, respectively, $L$ is the total round-trip length of the cavity, and $\mathrm{FSR}=2\pi\,c/L$ is the free spectral range. The last term in Eq.~(\ref{eq:modes}) describes a $\pi$ phase shift for antisymmetric tangential modes in a cavity with an odd number of mirrors \cite{Lange98}. For simplicity we have omitted the unknown net phase shift due to the dielectric mirror coatings, which leads to an offset of $\sim 1500~$MHz between s- and p-polarizations in our cavity.

In principle one can keep fixed either the probe laser frequency or the cavity length, and scan the other to determine the free spectral range (and therefore the cavity length). However the piezo scans of our laser and cavity are not linear enough over the required few-GHz range to accurately do this. Instead we match a total of 15 transverse modes, with splittings ranging from $9\,$--$1200$~MHz. The Pound-Drever-Hall sidebands provide a local frequency calibration. An ABCD matrix calculation is then performed using $L$ as a free parameter to match the observed splittings. The fitting is most tightly constrained by the resonance pairs with smallest splittings, but all of the splittings are consistent. We obtain $L=9.51(5)$~cm and $\mathrm{FSR}=2\pi\times3151(16)~$MHz. We have included the effect of a $0.5\%$ uncertainty stemming from the uncertainty on ROC as specified by the manufacturer. This value of $L$ is a percent or two smaller than the design length, but we do observe that the cavity mode is not perfectly centred on the mirrors. Given this value of FSR, we calculate a finesse of $\mathcal{F}=1710(60)$. Knowing $L$ we can also infer the cavity mode spot size, and thus the Rabi frequency $2g$ between a single atom and photon. In everything that follows, we restrict ourselves to the TEM$_{00}$ spatial mode of the cavity. The calculated $1/e^2$ intensity radii are $w_\parallel=90.2(5)~\mu$m and $w_\perp=128.0(3)~\mu$m. The electric dipole moment for the D$_1$ transitions (wavelength $\lambda=770.1$~nm \cite{Grosche06} and natural atomic linewidth $\gamma=2\pi\times2.978(6)$~MHz \cite{Stwalley97}) is $d=[3\epsilon_0\hbar\gamma\lambda^3/(4\pi^2)]^{1/2}=2.905\,ea_0$ (here $e$ is the electron charge and $a_0$ is the Bohr radius). Then $g=[d^2\omega_c/(2\hbar\epsilon_0V)]^{1/2}=2\pi\times91.5(5)$~kHz, where $V=\int d\mathbf{x}\,|\mathcal{E}(\mathbf{x})|^2=2.40(3)~\mathrm{mm}^3$ is the cavity mode volume for a peak-normalized field mode function $\mathcal{E}(\mathbf{x})$, and $\omega_c$ is the cavity resonance frequency.

\section{Collective strong coupling\label{sec:science}}

As discussed above, collective strong coupling between the cavity field and the atomic ensemble is evidenced by the normal mode or vacuum Rabi splitting of the cavity transmission spectrum. Given a number density of atoms $\varrho(\mathbf{x})$, the effective number of atoms in the cavity mode is $N=\int d\mathbf{x}\,\varrho(\mathbf{x})|\mathcal{E}(\mathbf{x})|^2$  \cite{Carmichael99} and the vacuum Rabi frequency becomes $G=g\,(\xi N)^{1/2}$ \cite{Agarwal84}. The factor $\xi=5/18$ is the relative oscillator strength averaged over all of the $F=1\leftrightarrow F'=2$ transitions. Our cloud has an approximately spherical Gaussian density distribution, with a root-mean-squared size, $\sigma\sim 0.8$~mm, which is large compared to $w_\parallel$ and $w_\perp$ and small compared to the corresponding Rayleigh ranges. Then $N\approx(\pi^3/2)^{1/2}\varrho(0)\,\sigma\,w_\parallel w_\perp$, and a typical peak density of $10^9$~cm$^{-3}$ gives $N\approx 4\times 10^4$ and $G=2\pi\times 9$~MHz, which is well into the regime of collective strong coupling.

We begin each experimental run by collecting several million atoms in the 3D-MOT, and then blocking the pushing beam with the shutter. The repumping light is extinguished $100~\mu$s before the cooling light, in order to optically pump atoms into the $F=1$ ground states. The weak probe light (typically on the order of 1~nW before the cavity) and the magnetic field gradient are left on during the entire experimental cycle. The probe frequency is swept for $100~\mu$s and then the atoms are recaptured. Separate time-of-flight measurements yield a temperature of $\sim700~\mu$K, and show that the cloud expansion is negligible over the duration of the probe scan. The transmitted probe signal at the APD is recorded and averaged on a digital oscilloscope.

\begin{figure}[ht]
\centering
\includegraphics[width=0.4\textwidth]{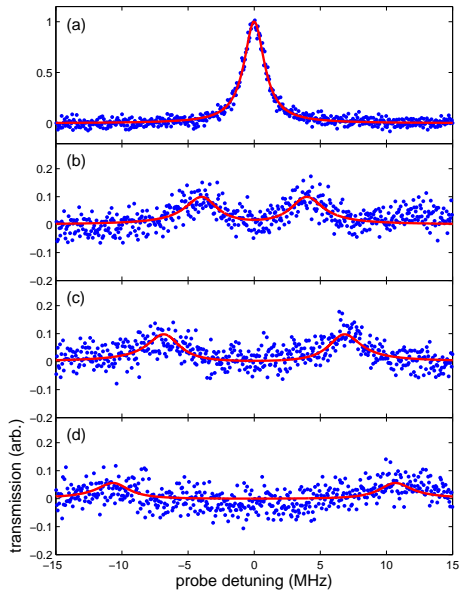}
\caption{Vacuum Rabi splitting for increasing atom number. (a) Empty cavity transmission spectrum for the TEM$_{00}$ mode, with $\Delta_a=\Delta_c$. Blue points are the data, averaged over 32 individual spectra (before storage), with $1~\mu$s sampling time. The red curve is a fit to a Lorentzian. In (b)--(d), the intracavity atom number is varied by displacing the MOT through the cavity mode. The fits to Eq.~(\ref{eq:T}) give $N=7.47(6)\times 10^3$, $2.04(2)\times 10^4$, and $4.9(1.5)\times 10^4$, respectively.}\label{fig4}
\end{figure}

\begin{figure}[ht]
\centering
\includegraphics[width=0.7\textwidth]{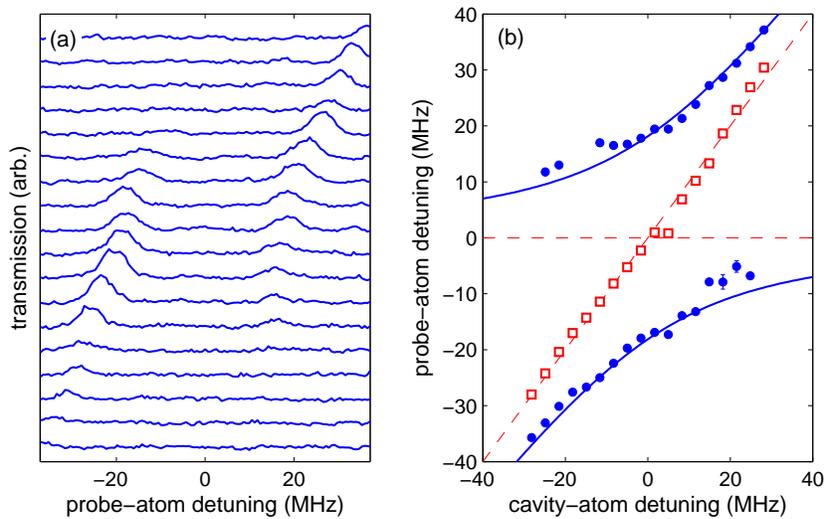}
\caption{Avoided crossing. (a) Transmission spectra for fixed $N$ and with cavity-atom detuning increasing vertically. The spectra have been offset for clarity. (b) Normal mode resonance frequencies, obtained through Lorentzian fits to the peaks in (a). The blue circles show the data, and the error bars show the standard errors from the fits. The solid blue curves show the prediction of Eq.~(\ref{eq:xing}) with $G=2\pi\times 18.1$~MHz. Red squares show data without atoms, and the horizontal (diagonal) dashed line shows the uncoupled atomic (cavity) resonance frequency.}\label{fig5}
\end{figure}

Example transmission spectra are shown in Fig.~\ref{fig4}, for the case where the cavity is on resonance with the free-space atomic transition. The probe power was $<1$~nW before the cavity. The intra-cavity atom number $N$ was varied by translating the centre of the MOT through the cavity mode using an added uniform magnetic field. The transmission spectra are well described by the CQED prediction \cite{Agarwal84},
\begin{eqnarray}\label{eq:T}
T &=& \frac{\kappa^2}{\left|(\kappa-i\Delta_c)+G^2/(\gamma-i\Delta_a)\right|^2} \quad.
\end{eqnarray}
Here $\Delta_c$ is the detuning between the probe laser and the uncoupled cavity, and $\Delta_a$ is the probe-atom detuning, which are equal for the data in Fig.~\ref{fig4}. Equation (\ref{eq:T}) assumes that the atomic excited-state population is negligible. With atoms in the cavity, the normal-mode splitting is apparent; as $G$ is increased, the resonance frequencies approach $\pm G$, the widths approach $(\kappa+\gamma)/2$, and the amplitudes approach $(1+\gamma/\kappa)^{-2}$. Fits to Eq.~(\ref{eq:T}) allow us to determine $N=7.47(6)\times 10^3$, $2.04(2)\times 10^4$, and $4.9(1.5)\times 10^4$ in panels (b)--(d), respectively. Independent in-situ fluorescence images of the MOT imply a maximum value of $N=3\times 10^4$. We expect the images to underestimate the atom number, since we conservatively overestimate the solid angle of the collected light by using the full clear aperture of the imaging lens.

When the cavity is detuned from resonance with the uncoupled atomic transition, the atoms induce a dispersive shift to the cavity resonance in addition to the splitting just described \cite{CavsClassical}. By taking two-dimensional scans over $\Delta_c$ and $\Delta_a$, it is possible to map out the avoided crossing of the normal modes induced by the coupling \cite{Orozco97}. This is shown in Fig.~\ref{fig5} for larger MOT number. In (a) we show the cavity transmission spectra, with the cavity-atom detuning ($\Delta_a-\Delta_c$) increasing vertically. Note that at large probe detunings, the incident probe power is reduced due to the finite bandwidth of the AOM. This will be compensated in future experiments with an active feedback system. Here we are not concerned with the peak heights, and simply normalize all of the traces to the maximum incident power. When we track the peak positions, we clearly see the avoided crossing, as shown in Fig.~\ref{fig5}(b). The data are again well described by the theory in \cite{Agarwal84}, which gives the normal mode resonance frequencies,
\begin{eqnarray}\label{eq:xing}
\Delta_a^\pm &=& \frac{\Delta_a-\Delta_c}{2} \pm\sqrt{G^2+\left(\frac{\Delta_a-\Delta_c}{2}\right)^{\!\!2}} \quad.
\end{eqnarray}
For these data $G=2\pi\times 18.1(7)~$MHz, implying $N=1.01(8)\times 10^5$. The observed splitting corresponds to a collective cooperativity of $C=G^2/(\kappa\gamma)=119(9)$. The cooperativity is the central parameter describing the dominance of the atomic coupling with the cavity mode over the continuum of free-space modes, as well as the onset of optical nonlinearities \cite{YeCavs,CavsClassical,Orozco97}.

\section{Outlook\label{sec:outlook}}

We have described an apparatus for studying ensemble cavity QED in the regime of collective strong coupling. Potassium-39 atoms are cooled in a 2D-MOT and transferred to a 3D-MOT overlapping the mode of a high-finesse ring cavity. We have characterized the properties of the cavity which are relevant to understanding the atom-light coupling. We have demonstrated collective strong coupling through observations of the vacuum Rabi splitting of the cavity transmission spectrum for varying numbers of atoms. Finally, we have observed the avoided crossing of the normal modes of the coupled system.

We next aim to control the group index and optical gain of the atomic medium. It is well known that electromagnetically induced transparency (EIT) can lead to large refractive group indices \cite{Scully91}. We will apply the laser system described in \cite{Lampis16} to our potassium MOT. We can estimate the group index of the intracavity EIT medium as $n_g\sim(2G/\Gamma)^2$, where $\Gamma$ is the EIT linewidth. For our current conditions, Doppler broadening of the two-photon transition limits $\Gamma\sim2\pi\times 0.6$~MHz, but standard methods could reduce the MOT temperature to $\sim30~\mu$K \cite{Mumbai11,Modugno11,Marcassa12}, for which $\Gamma\sim2\pi\times 0.1$~MHz. At that level the magnetic field variations due to the MOT gradient over the size of the atom-light overlap region will dominate, giving $\Gamma\sim2\pi\times 0.3$~MHz. This implies a group index of several $10^4$, allowing wide-ranging control over the light scattering dynamics in the cavity \cite{Laupretre11}, with minimal absorption losses, and in the strong coupling regime. We can also study lasing with the cold potassium atoms as the gain medium \cite{Giacobino92,Kaiser08,Kasevich11,Thompson12,Thompson16}. For superradiant (slow-light) lasing, the group index is approximately equal to $\kappa/\mathrm{GBW}$ where GBW is the gain bandwidth \cite{Woerdman94,Thompson12}. In this case lasing on the lower-finesse p-polarized mode of the ring cavity would be advantageous. Finally we note that our ring geometry makes our system attractive for studying the dynamics of anomalous dispersion \cite{Laupretre12} as applied towards superluminal enhancement of rotation sensing in a ring laser gyro \cite{Shahriar07}.

\section*{Acknowledgments}
This work was funded by EPSRC (EP/J016985/1), DSTL (DSTLX1000092132), and the University of Birmingham in the UK, and by FAPESP in Brazil. M.H.~acknowledges additional support from the Birmingham Visiting Brazil Fellows program. We are grateful to Kai Bongs and Giovanni Barontini for the loan of equipment, and Stephen Brooks and the Birmingham mechanical workshop for expert technical assistance. All data appearing in this work are available online \cite{data}.

%\bibliographystyle{plain}
\bibliography{RingCav}

\begin{thebibliography}{51}%
\makeatletter
\providecommand \@ifxundefined [1]{%
 \@ifx{#1\undefined}
}%
\providecommand \@ifnum [1]{%
 \ifnum #1\expandafter \@firstoftwo
 \else \expandafter \@secondoftwo
 \fi
}%
\providecommand \@ifx [1]{%
 \ifx #1\expandafter \@firstoftwo
 \else \expandafter \@secondoftwo
 \fi
}%
\providecommand \natexlab [1]{#1}%
\providecommand \enquote  [1]{``#1''}%
\providecommand \bibnamefont  [1]{#1}%
\providecommand \bibfnamefont [1]{#1}%
\providecommand \citenamefont [1]{#1}%
\providecommand \href@noop [0]{\@secondoftwo}%
\providecommand \href [0]{\begingroup \@sanitize@url \@href}%
\providecommand \@href[1]{\@@startlink{#1}\@@href}%
\providecommand \@@href[1]{\endgroup#1\@@endlink}%
\providecommand \@sanitize@url [0]{\catcode `\\12\catcode `\$12\catcode
  `\&12\catcode `\#12\catcode `\^12\catcode `\_12\catcode `\%12\relax}%
\providecommand \@@startlink[1]{}%
\providecommand \@@endlink[0]{}%
\providecommand \url  [0]{\begingroup\@sanitize@url \@url }%
\providecommand \@url [1]{\endgroup\@href {#1}{\urlprefix }}%
\providecommand \urlprefix  [0]{URL }%
\providecommand \Eprint [0]{\href }%
\providecommand \doibase [0]{http://dx.doi.org/}%
\providecommand \selectlanguage [0]{\@gobble}%
\providecommand \bibinfo  [0]{\@secondoftwo}%
\providecommand \bibfield  [0]{\@secondoftwo}%
\providecommand \translation [1]{[#1]}%
\providecommand \BibitemOpen [0]{}%
\providecommand \bibitemStop [0]{}%
\providecommand \bibitemNoStop [0]{.\EOS\space}%
\providecommand \EOS [0]{\spacefactor3000\relax}%
\providecommand \BibitemShut  [1]{\csname bibitem#1\endcsname}%
\let\auto@bib@innerbib\@empty
%</preamble>
\bibitem [{\citenamefont {Jaynes}\ and\ \citenamefont
  {Cummings}(1963)}]{JaynesCummings}%
  \BibitemOpen
  \bibfield  {author} {\bibinfo {author} {\bibfnamefont {E.~T.}\ \bibnamefont
  {Jaynes}}\ and\ \bibinfo {author} {\bibfnamefont {F.~W.}\ \bibnamefont
  {Cummings}},\ }\href {\doibase 10.1109/PROC.1963.1664} {\bibfield  {journal}
  {\bibinfo  {journal} {Proceedings of the IEEE}\ }\textbf {\bibinfo {volume}
  {51}},\ \bibinfo {pages} {89} (\bibinfo {year} {1963})}\BibitemShut {NoStop}%
\bibitem [{\citenamefont {Ye}\ and\ \citenamefont {Lynn}(2003)}]{YeCavs}%
  \BibitemOpen
  \bibfield  {author} {\bibinfo {author} {\bibfnamefont {J.}~\bibnamefont
  {Ye}}\ and\ \bibinfo {author} {\bibfnamefont {T.~W.}\ \bibnamefont {Lynn}},\
  }in\ \href {\doibase 10.1016/S1049-250X(03)80003-4} {\emph {\bibinfo
  {booktitle} {Advances in Atomic, Molecular, and Optical Physics}}},\ \bibinfo
  {series} {Advances In Atomic, Molecular, and Optical Physics}, Vol.~\bibinfo
  {volume} {49},\ \bibinfo {editor} {edited by\ \bibinfo {editor}
  {\bibfnamefont {B.}~\bibnamefont {Bederson}}\ and\ \bibinfo {editor}
  {\bibfnamefont {H.}~\bibnamefont {Walther}}}\ (\bibinfo  {publisher}
  {Academic Press},\ \bibinfo {year} {2003})\ pp.\ \bibinfo {pages} {1 --
  83}\BibitemShut {NoStop}%
\bibitem [{\citenamefont {Tanji-Suzuki}\ \emph {et~al.}(2011)\citenamefont
  {Tanji-Suzuki}, \citenamefont {Leroux}, \citenamefont {Schleier-Smith},
  \citenamefont {Cetina}, \citenamefont {Grier}, \citenamefont {Simon},\ and\
  \citenamefont {Vuletic}}]{CavsClassical}%
  \BibitemOpen
  \bibfield  {author} {\bibinfo {author} {\bibfnamefont {H.}~\bibnamefont
  {Tanji-Suzuki}}, \bibinfo {author} {\bibfnamefont {I.~D.}\ \bibnamefont
  {Leroux}}, \bibinfo {author} {\bibfnamefont {M.~H.}\ \bibnamefont
  {Schleier-Smith}}, \bibinfo {author} {\bibfnamefont {M.}~\bibnamefont
  {Cetina}}, \bibinfo {author} {\bibfnamefont {A.~T.}\ \bibnamefont {Grier}},
  \bibinfo {author} {\bibfnamefont {J.}~\bibnamefont {Simon}}, \ and\ \bibinfo
  {author} {\bibfnamefont {V.}~\bibnamefont {Vuletic}},\ }in\ \href {\doibase
  10.1016/B978-0-12-385508-4.00004-8} {\emph {\bibinfo {booktitle} {Advances in
  Atomic, Molecular, and Optical Physics}}},\ \bibinfo {series} {Advances In
  Atomic, Molecular, and Optical Physics}, Vol.~\bibinfo {volume} {60},\
  \bibinfo {editor} {edited by\ \bibinfo {editor} {\bibfnamefont {P.~B.}\
  \bibnamefont {E.~Arimondo}}\ and\ \bibinfo {editor} {\bibfnamefont
  {C.}~\bibnamefont {Lin}}}\ (\bibinfo  {publisher} {Academic Press},\ \bibinfo
  {year} {2011})\ pp.\ \bibinfo {pages} {201 -- 237}\BibitemShut {NoStop}%
\bibitem [{\citenamefont {Sanchez-Mondragon}\ \emph {et~al.}(1983)\citenamefont
  {Sanchez-Mondragon}, \citenamefont {Narozhny},\ and\ \citenamefont
  {Eberly}}]{Eberly83}%
  \BibitemOpen
  \bibfield  {author} {\bibinfo {author} {\bibfnamefont {J.~J.}\ \bibnamefont
  {Sanchez-Mondragon}}, \bibinfo {author} {\bibfnamefont {N.~B.}\ \bibnamefont
  {Narozhny}}, \ and\ \bibinfo {author} {\bibfnamefont {J.~H.}\ \bibnamefont
  {Eberly}},\ }\href {\doibase 10.1103/PhysRevLett.51.550} {\bibfield
  {journal} {\bibinfo  {journal} {Phys. Rev. Lett.}\ }\textbf {\bibinfo
  {volume} {51}},\ \bibinfo {pages} {550} (\bibinfo {year} {1983})}\BibitemShut
  {NoStop}%
\bibitem [{\citenamefont {Raizen}\ \emph {et~al.}(1989)\citenamefont {Raizen},
  \citenamefont {Thompson}, \citenamefont {Brecha}, \citenamefont {Kimble},\
  and\ \citenamefont {Carmichael}}]{Raizen89}%
  \BibitemOpen
  \bibfield  {author} {\bibinfo {author} {\bibfnamefont {M.~G.}\ \bibnamefont
  {Raizen}}, \bibinfo {author} {\bibfnamefont {R.~J.}\ \bibnamefont
  {Thompson}}, \bibinfo {author} {\bibfnamefont {R.~J.}\ \bibnamefont
  {Brecha}}, \bibinfo {author} {\bibfnamefont {H.~J.}\ \bibnamefont {Kimble}},
  \ and\ \bibinfo {author} {\bibfnamefont {H.~J.}\ \bibnamefont {Carmichael}},\
  }\href {\doibase 10.1103/PhysRevLett.63.240} {\bibfield  {journal} {\bibinfo
  {journal} {Phys. Rev. Lett.}\ }\textbf {\bibinfo {volume} {63}},\ \bibinfo
  {pages} {240} (\bibinfo {year} {1989})}\BibitemShut {NoStop}%
\bibitem [{\citenamefont {Zhu}\ \emph {et~al.}(1990)\citenamefont {Zhu},
  \citenamefont {Gauthier}, \citenamefont {Morin}, \citenamefont {Wu},
  \citenamefont {Carmichael},\ and\ \citenamefont {Mossberg}}]{Mossberg90}%
  \BibitemOpen
  \bibfield  {author} {\bibinfo {author} {\bibfnamefont {Y.}~\bibnamefont
  {Zhu}}, \bibinfo {author} {\bibfnamefont {D.~J.}\ \bibnamefont {Gauthier}},
  \bibinfo {author} {\bibfnamefont {S.~E.}\ \bibnamefont {Morin}}, \bibinfo
  {author} {\bibfnamefont {Q.}~\bibnamefont {Wu}}, \bibinfo {author}
  {\bibfnamefont {H.~J.}\ \bibnamefont {Carmichael}}, \ and\ \bibinfo {author}
  {\bibfnamefont {T.~W.}\ \bibnamefont {Mossberg}},\ }\href {\doibase
  10.1103/PhysRevLett.64.2499} {\bibfield  {journal} {\bibinfo  {journal}
  {Phys. Rev. Lett.}\ }\textbf {\bibinfo {volume} {64}},\ \bibinfo {pages}
  {2499} (\bibinfo {year} {1990})}\BibitemShut {NoStop}%
\bibitem [{\citenamefont {Thompson}\ \emph {et~al.}(1992)\citenamefont
  {Thompson}, \citenamefont {Rempe},\ and\ \citenamefont {Kimble}}]{TRK92}%
  \BibitemOpen
  \bibfield  {author} {\bibinfo {author} {\bibfnamefont {R.~J.}\ \bibnamefont
  {Thompson}}, \bibinfo {author} {\bibfnamefont {G.}~\bibnamefont {Rempe}}, \
  and\ \bibinfo {author} {\bibfnamefont {H.~J.}\ \bibnamefont {Kimble}},\
  }\href {\doibase 10.1103/PhysRevLett.68.1132} {\bibfield  {journal} {\bibinfo
   {journal} {Phys. Rev. Lett.}\ }\textbf {\bibinfo {volume} {68}},\ \bibinfo
  {pages} {1132} (\bibinfo {year} {1992})}\BibitemShut {NoStop}%
\bibitem [{\citenamefont {Tavis}\ and\ \citenamefont
  {Cummings}(1968)}]{TavisCummings}%
  \BibitemOpen
  \bibfield  {author} {\bibinfo {author} {\bibfnamefont {M.}~\bibnamefont
  {Tavis}}\ and\ \bibinfo {author} {\bibfnamefont {F.~W.}\ \bibnamefont
  {Cummings}},\ }\href {\doibase 10.1103/PhysRev.170.379} {\bibfield  {journal}
  {\bibinfo  {journal} {Phys. Rev.}\ }\textbf {\bibinfo {volume} {170}},\
  \bibinfo {pages} {379} (\bibinfo {year} {1968})}\BibitemShut {NoStop}%
\bibitem [{\citenamefont {Agarwal}(1984)}]{Agarwal84}%
  \BibitemOpen
  \bibfield  {author} {\bibinfo {author} {\bibfnamefont {G.~S.}\ \bibnamefont
  {Agarwal}},\ }\href {\doibase 10.1103/PhysRevLett.53.1732} {\bibfield
  {journal} {\bibinfo  {journal} {Phys. Rev. Lett.}\ }\textbf {\bibinfo
  {volume} {53}},\ \bibinfo {pages} {1732} (\bibinfo {year}
  {1984})}\BibitemShut {NoStop}%
\bibitem [{\citenamefont {von Cube}\ \emph {et~al.}(2006)\citenamefont {von
  Cube}, \citenamefont {Slama}, \citenamefont {Kohler}, \citenamefont
  {Zimmermann},\ and\ \citenamefont {Courteille}}]{Courteille06}%
  \BibitemOpen
  \bibfield  {author} {\bibinfo {author} {\bibfnamefont {C.}~\bibnamefont {von
  Cube}}, \bibinfo {author} {\bibfnamefont {S.}~\bibnamefont {Slama}}, \bibinfo
  {author} {\bibfnamefont {M.}~\bibnamefont {Kohler}}, \bibinfo {author}
  {\bibfnamefont {C.}~\bibnamefont {Zimmermann}}, \ and\ \bibinfo {author}
  {\bibfnamefont {P.}~\bibnamefont {Courteille}},\ }\href {\doibase
  10.1002/prop.200610307} {\bibfield  {journal} {\bibinfo  {journal}
  {Fortschritte der Physik}\ }\textbf {\bibinfo {volume} {54}},\ \bibinfo
  {pages} {726} (\bibinfo {year} {2006})}\BibitemShut {NoStop}%
\bibitem [{\citenamefont {Ritsch}\ \emph {et~al.}(2013)\citenamefont {Ritsch},
  \citenamefont {Domokos}, \citenamefont {Brennecke},\ and\ \citenamefont
  {Esslinger}}]{RitschRMP}%
  \BibitemOpen
  \bibfield  {author} {\bibinfo {author} {\bibfnamefont {H.}~\bibnamefont
  {Ritsch}}, \bibinfo {author} {\bibfnamefont {P.}~\bibnamefont {Domokos}},
  \bibinfo {author} {\bibfnamefont {F.}~\bibnamefont {Brennecke}}, \ and\
  \bibinfo {author} {\bibfnamefont {T.}~\bibnamefont {Esslinger}},\ }\href
  {\doibase 10.1103/RevModPhys.85.553} {\bibfield  {journal} {\bibinfo
  {journal} {Rev. Mod. Phys.}\ }\textbf {\bibinfo {volume} {85}},\ \bibinfo
  {pages} {553} (\bibinfo {year} {2013})}\BibitemShut {NoStop}%
\bibitem [{\citenamefont {Klinner}\ \emph {et~al.}(2006)\citenamefont
  {Klinner}, \citenamefont {Lindholdt}, \citenamefont {Nagorny},\ and\
  \citenamefont {Hemmerich}}]{Hemmerich06}%
  \BibitemOpen
  \bibfield  {author} {\bibinfo {author} {\bibfnamefont {J.}~\bibnamefont
  {Klinner}}, \bibinfo {author} {\bibfnamefont {M.}~\bibnamefont {Lindholdt}},
  \bibinfo {author} {\bibfnamefont {B.}~\bibnamefont {Nagorny}}, \ and\
  \bibinfo {author} {\bibfnamefont {A.}~\bibnamefont {Hemmerich}},\ }\href
  {\doibase 10.1103/PhysRevLett.96.023002} {\bibfield  {journal} {\bibinfo
  {journal} {Phys. Rev. Lett.}\ }\textbf {\bibinfo {volume} {96}},\ \bibinfo
  {pages} {023002} (\bibinfo {year} {2006})}\BibitemShut {NoStop}%
\bibitem [{\citenamefont {Tuchman}\ \emph {et~al.}(2006)\citenamefont
  {Tuchman}, \citenamefont {Long}, \citenamefont {Vrijsen}, \citenamefont
  {Boudet}, \citenamefont {Lee},\ and\ \citenamefont {Kasevich}}]{Kasevich06}%
  \BibitemOpen
  \bibfield  {author} {\bibinfo {author} {\bibfnamefont {A.~K.}\ \bibnamefont
  {Tuchman}}, \bibinfo {author} {\bibfnamefont {R.}~\bibnamefont {Long}},
  \bibinfo {author} {\bibfnamefont {G.}~\bibnamefont {Vrijsen}}, \bibinfo
  {author} {\bibfnamefont {J.}~\bibnamefont {Boudet}}, \bibinfo {author}
  {\bibfnamefont {J.}~\bibnamefont {Lee}}, \ and\ \bibinfo {author}
  {\bibfnamefont {M.~A.}\ \bibnamefont {Kasevich}},\ }\href {\doibase
  10.1103/PhysRevA.74.053821} {\bibfield  {journal} {\bibinfo  {journal} {Phys.
  Rev. A}\ }\textbf {\bibinfo {volume} {74}},\ \bibinfo {pages} {053821}
  (\bibinfo {year} {2006})}\BibitemShut {NoStop}%
\bibitem [{\citenamefont {Chen}\ \emph {et~al.}(2011)\citenamefont {Chen},
  \citenamefont {Bohnet}, \citenamefont {Sankar}, \citenamefont {Dai},\ and\
  \citenamefont {Thompson}}]{Thompson11}%
  \BibitemOpen
  \bibfield  {author} {\bibinfo {author} {\bibfnamefont {Z.}~\bibnamefont
  {Chen}}, \bibinfo {author} {\bibfnamefont {J.~G.}\ \bibnamefont {Bohnet}},
  \bibinfo {author} {\bibfnamefont {S.~R.}\ \bibnamefont {Sankar}}, \bibinfo
  {author} {\bibfnamefont {J.}~\bibnamefont {Dai}}, \ and\ \bibinfo {author}
  {\bibfnamefont {J.~K.}\ \bibnamefont {Thompson}},\ }\href {\doibase
  10.1103/PhysRevLett.106.133601} {\bibfield  {journal} {\bibinfo  {journal}
  {Phys. Rev. Lett.}\ }\textbf {\bibinfo {volume} {106}},\ \bibinfo {pages}
  {133601} (\bibinfo {year} {2011})}\BibitemShut {NoStop}%
\bibitem [{\citenamefont {Zhang}\ \emph {et~al.}(2010)\citenamefont {Zhang},
  \citenamefont {Wei}, \citenamefont {Hernandez},\ and\ \citenamefont
  {Zhu}}]{Zhu10}%
  \BibitemOpen
  \bibfield  {author} {\bibinfo {author} {\bibfnamefont {J.}~\bibnamefont
  {Zhang}}, \bibinfo {author} {\bibfnamefont {X.}~\bibnamefont {Wei}}, \bibinfo
  {author} {\bibfnamefont {G.}~\bibnamefont {Hernandez}}, \ and\ \bibinfo
  {author} {\bibfnamefont {Y.}~\bibnamefont {Zhu}},\ }\href {\doibase
  10.1103/PhysRevA.81.033804} {\bibfield  {journal} {\bibinfo  {journal} {Phys.
  Rev. A}\ }\textbf {\bibinfo {volume} {81}},\ \bibinfo {pages} {033804}
  (\bibinfo {year} {2010})}\BibitemShut {NoStop}%
\bibitem [{\citenamefont {Arnold}\ \emph {et~al.}(2011)\citenamefont {Arnold},
  \citenamefont {Baden},\ and\ \citenamefont {Barrett}}]{Barrett11}%
  \BibitemOpen
  \bibfield  {author} {\bibinfo {author} {\bibfnamefont {K.~J.}\ \bibnamefont
  {Arnold}}, \bibinfo {author} {\bibfnamefont {M.~P.}\ \bibnamefont {Baden}}, \
  and\ \bibinfo {author} {\bibfnamefont {M.~D.}\ \bibnamefont {Barrett}},\
  }\href {\doibase 10.1103/PhysRevA.84.033843} {\bibfield  {journal} {\bibinfo
  {journal} {Phys. Rev. A}\ }\textbf {\bibinfo {volume} {84}},\ \bibinfo
  {pages} {033843} (\bibinfo {year} {2011})}\BibitemShut {NoStop}%
\bibitem [{\citenamefont {Wickenbrock}\ \emph {et~al.}(2013)\citenamefont
  {Wickenbrock}, \citenamefont {Hemmerling}, \citenamefont {Robb},
  \citenamefont {Emary},\ and\ \citenamefont {Renzoni}}]{Renzoni13}%
  \BibitemOpen
  \bibfield  {author} {\bibinfo {author} {\bibfnamefont {A.}~\bibnamefont
  {Wickenbrock}}, \bibinfo {author} {\bibfnamefont {M.}~\bibnamefont
  {Hemmerling}}, \bibinfo {author} {\bibfnamefont {G.~R.~M.}\ \bibnamefont
  {Robb}}, \bibinfo {author} {\bibfnamefont {C.}~\bibnamefont {Emary}}, \ and\
  \bibinfo {author} {\bibfnamefont {F.}~\bibnamefont {Renzoni}},\ }\href
  {\doibase 10.1103/PhysRevA.87.043817} {\bibfield  {journal} {\bibinfo
  {journal} {Phys. Rev. A}\ }\textbf {\bibinfo {volume} {87}},\ \bibinfo
  {pages} {043817} (\bibinfo {year} {2013})}\BibitemShut {NoStop}%
\bibitem [{\citenamefont {Ningyuan}\ \emph {et~al.}(2016)\citenamefont
  {Ningyuan}, \citenamefont {Georgakopoulos}, \citenamefont {Ryou},
  \citenamefont {Schine}, \citenamefont {Sommer},\ and\ \citenamefont
  {Simon}}]{Simon16}%
  \BibitemOpen
  \bibfield  {author} {\bibinfo {author} {\bibfnamefont {J.}~\bibnamefont
  {Ningyuan}}, \bibinfo {author} {\bibfnamefont {A.}~\bibnamefont
  {Georgakopoulos}}, \bibinfo {author} {\bibfnamefont {A.}~\bibnamefont
  {Ryou}}, \bibinfo {author} {\bibfnamefont {N.}~\bibnamefont {Schine}},
  \bibinfo {author} {\bibfnamefont {A.}~\bibnamefont {Sommer}}, \ and\ \bibinfo
  {author} {\bibfnamefont {J.}~\bibnamefont {Simon}},\ }\href {\doibase
  10.1103/PhysRevA.93.041802} {\bibfield  {journal} {\bibinfo  {journal} {Phys.
  Rev. A}\ }\textbf {\bibinfo {volume} {93}},\ \bibinfo {pages} {041802}
  (\bibinfo {year} {2016})}\BibitemShut {NoStop}%
\bibitem [{\citenamefont {Chin}\ \emph {et~al.}(2010)\citenamefont {Chin},
  \citenamefont {Grimm}, \citenamefont {Julienne},\ and\ \citenamefont
  {Tiesinga}}]{FeshbachRMP}%
  \BibitemOpen
  \bibfield  {author} {\bibinfo {author} {\bibfnamefont {C.}~\bibnamefont
  {Chin}}, \bibinfo {author} {\bibfnamefont {R.}~\bibnamefont {Grimm}},
  \bibinfo {author} {\bibfnamefont {P.}~\bibnamefont {Julienne}}, \ and\
  \bibinfo {author} {\bibfnamefont {E.}~\bibnamefont {Tiesinga}},\ }\href
  {\doibase 10.1103/RevModPhys.82.1225} {\bibfield  {journal} {\bibinfo
  {journal} {Rev. Mod. Phys.}\ }\textbf {\bibinfo {volume} {82}},\ \bibinfo
  {pages} {1225} (\bibinfo {year} {2010})}\BibitemShut {NoStop}%
\bibitem [{\citenamefont {Birnbaum}\ \emph {et~al.}(2006)\citenamefont
  {Birnbaum}, \citenamefont {Parkins},\ and\ \citenamefont
  {Kimble}}]{Birnbaum06}%
  \BibitemOpen
  \bibfield  {author} {\bibinfo {author} {\bibfnamefont {K.~M.}\ \bibnamefont
  {Birnbaum}}, \bibinfo {author} {\bibfnamefont {A.~S.}\ \bibnamefont
  {Parkins}}, \ and\ \bibinfo {author} {\bibfnamefont {H.~J.}\ \bibnamefont
  {Kimble}},\ }\href {\doibase 10.1103/PhysRevA.74.063802} {\bibfield
  {journal} {\bibinfo  {journal} {Phys. Rev. A}\ }\textbf {\bibinfo {volume}
  {74}},\ \bibinfo {pages} {063802} (\bibinfo {year} {2006})}\BibitemShut
  {NoStop}%
\bibitem [{Pro()}]{Products}%
  \BibitemOpen
  \href@noop {} {}\bibinfo {note} {Trade names and product numbers are used for
  identification purposes only, and do not constitute an endorsement by the
  authors or their institutions.}\BibitemShut {Stop}%
\bibitem [{\citenamefont {Pahwa}(2014)}]{ThesisPahwa}%
  \BibitemOpen
  \bibfield  {author} {\bibinfo {author} {\bibfnamefont {K.}~\bibnamefont
  {Pahwa}},\ }\emph {\bibinfo {title} {Magneto optical trapping of potassium-39
  in a ring cavity}},\ \href {http://etheses.bham.ac.uk/5533/} {Ph.D. thesis},\
  \bibinfo  {school} {University of Birmingham} (\bibinfo {year}
  {2014})\BibitemShut {NoStop}%
\bibitem [{\citenamefont {Mudarikwa}(2015)}]{ThesisMudarikwa}%
  \BibitemOpen
  \bibfield  {author} {\bibinfo {author} {\bibfnamefont {L.}~\bibnamefont
  {Mudarikwa}},\ }\emph {\bibinfo {title} {Cold atoms in a ring cavity}},\
  \href {http://etheses.bham.ac.uk/5843/} {Ph.D. thesis},\ \bibinfo  {school}
  {University of Birmingham} (\bibinfo {year} {2015})\BibitemShut {NoStop}%
\bibitem [{\citenamefont {Mudarikwa}\ \emph {et~al.}(2012)\citenamefont
  {Mudarikwa}, \citenamefont {Pahwa},\ and\ \citenamefont
  {Goldwin}}]{Mudarikwa12}%
  \BibitemOpen
  \bibfield  {author} {\bibinfo {author} {\bibfnamefont {L.}~\bibnamefont
  {Mudarikwa}}, \bibinfo {author} {\bibfnamefont {K.}~\bibnamefont {Pahwa}}, \
  and\ \bibinfo {author} {\bibfnamefont {J.}~\bibnamefont {Goldwin}},\ }\href
  {http://stacks.iop.org/0953-4075/45/i=6/a=065002} {\bibfield  {journal}
  {\bibinfo  {journal} {Journal of Physics B: Atomic, Molecular and Optical
  Physics}\ }\textbf {\bibinfo {volume} {45}},\ \bibinfo {pages} {065002}
  (\bibinfo {year} {2012})}\BibitemShut {NoStop}%
\bibitem [{\citenamefont {Pahwa}\ \emph {et~al.}(2012)\citenamefont {Pahwa},
  \citenamefont {Mudarikwa},\ and\ \citenamefont {Goldwin}}]{Pahwa12}%
  \BibitemOpen
  \bibfield  {author} {\bibinfo {author} {\bibfnamefont {K.}~\bibnamefont
  {Pahwa}}, \bibinfo {author} {\bibfnamefont {L.}~\bibnamefont {Mudarikwa}}, \
  and\ \bibinfo {author} {\bibfnamefont {J.}~\bibnamefont {Goldwin}},\ }\href
  {\doibase 10.1364/OE.20.017456} {\bibfield  {journal} {\bibinfo  {journal}
  {Opt. Express}\ }\textbf {\bibinfo {volume} {20}},\ \bibinfo {pages} {17456}
  (\bibinfo {year} {2012})}\BibitemShut {NoStop}%
\bibitem [{\citenamefont {Ritt}\ \emph {et~al.}(2004)\citenamefont {Ritt},
  \citenamefont {Cennini}, \citenamefont {Geckeler},\ and\ \citenamefont
  {Weitz}}]{Ritt04}%
  \BibitemOpen
  \bibfield  {author} {\bibinfo {author} {\bibfnamefont {G.}~\bibnamefont
  {Ritt}}, \bibinfo {author} {\bibfnamefont {G.}~\bibnamefont {Cennini}},
  \bibinfo {author} {\bibfnamefont {C.}~\bibnamefont {Geckeler}}, \ and\
  \bibinfo {author} {\bibfnamefont {M.}~\bibnamefont {Weitz}},\ }\href
  {\doibase 10.1007/s00340-004-1559-6} {\bibfield  {journal} {\bibinfo
  {journal} {Applied Physics B}\ }\textbf {\bibinfo {volume} {79}},\ \bibinfo
  {pages} {363} (\bibinfo {year} {2004})}\BibitemShut {NoStop}%
\bibitem [{Pot()}]{Potassium}%
  \BibitemOpen
  \href@noop {} {}\bibinfo {note} {As with all potassium-39 cooling
  experiments, the small excited-state hyperfine splittings are not well
  resolved in our system, so that each wavelength provides a cooling/trapping
  force on multiple transitions, with the assignments of $F$ and $F'$ and
  designations \textit{cooling} and \textit{repump} purely
  conventional.}\BibitemShut {Stop}%
\bibitem [{\citenamefont {Budker}\ \emph {et~al.}(2000)\citenamefont {Budker},
  \citenamefont {Rochester},\ and\ \citenamefont {Yashchuk}}]{Budker00}%
  \BibitemOpen
  \bibfield  {author} {\bibinfo {author} {\bibfnamefont {D.}~\bibnamefont
  {Budker}}, \bibinfo {author} {\bibfnamefont {S.~M.}\ \bibnamefont
  {Rochester}}, \ and\ \bibinfo {author} {\bibfnamefont {V.~V.}\ \bibnamefont
  {Yashchuk}},\ }\href {\doibase http://dx.doi.org/10.1063/1.1304879}
  {\bibfield  {journal} {\bibinfo  {journal} {Review of Scientific
  Instruments}\ }\textbf {\bibinfo {volume} {71}},\ \bibinfo {pages} {2984}
  (\bibinfo {year} {2000})}\BibitemShut {NoStop}%
\bibitem [{\citenamefont {Kruse}\ \emph {et~al.}(2003)\citenamefont {Kruse},
  \citenamefont {von Cube}, \citenamefont {Zimmermann},\ and\ \citenamefont
  {Courteille}}]{CARL}%
  \BibitemOpen
  \bibfield  {author} {\bibinfo {author} {\bibfnamefont {D.}~\bibnamefont
  {Kruse}}, \bibinfo {author} {\bibfnamefont {C.}~\bibnamefont {von Cube}},
  \bibinfo {author} {\bibfnamefont {C.}~\bibnamefont {Zimmermann}}, \ and\
  \bibinfo {author} {\bibfnamefont {P.~W.}\ \bibnamefont {Courteille}},\ }\href
  {\doibase 10.1103/PhysRevLett.91.183601} {\bibfield  {journal} {\bibinfo
  {journal} {Phys. Rev. Lett.}\ }\textbf {\bibinfo {volume} {91}},\ \bibinfo
  {pages} {183601} (\bibinfo {year} {2003})}\BibitemShut {NoStop}%
\bibitem [{\citenamefont {Bao}\ \emph {et~al.}(2012)\citenamefont {Bao},
  \citenamefont {Reingruber}, \citenamefont {Dietrich}, \citenamefont {Rui},
  \citenamefont {D\"{u}ck}, \citenamefont {Strassel}, \citenamefont {Li},
  \citenamefont {Liu}, \citenamefont {Zhao},\ and\ \citenamefont
  {Pan}}]{Pan12}%
  \BibitemOpen
  \bibfield  {author} {\bibinfo {author} {\bibfnamefont {X.-H.}\ \bibnamefont
  {Bao}}, \bibinfo {author} {\bibfnamefont {A.}~\bibnamefont {Reingruber}},
  \bibinfo {author} {\bibfnamefont {P.}~\bibnamefont {Dietrich}}, \bibinfo
  {author} {\bibfnamefont {J.}~\bibnamefont {Rui}}, \bibinfo {author}
  {\bibfnamefont {A.}~\bibnamefont {D\"{u}ck}}, \bibinfo {author}
  {\bibfnamefont {T.}~\bibnamefont {Strassel}}, \bibinfo {author}
  {\bibfnamefont {L.}~\bibnamefont {Li}}, \bibinfo {author} {\bibfnamefont
  {N.-L.}\ \bibnamefont {Liu}}, \bibinfo {author} {\bibfnamefont
  {B.}~\bibnamefont {Zhao}}, \ and\ \bibinfo {author} {\bibfnamefont {J.-W.}\
  \bibnamefont {Pan}},\ }\href {\doibase 10.1038/nphys2324} {\bibfield
  {journal} {\bibinfo  {journal} {Nature Physics}\ }\textbf {\bibinfo {volume}
  {8}},\ \bibinfo {pages} {517} (\bibinfo {year} {2012})}\BibitemShut {NoStop}%
\bibitem [{\citenamefont {Bernon}\ \emph {et~al.}(2011)\citenamefont {Bernon},
  \citenamefont {Vanderbruggen}, \citenamefont {Kohlhaas}, \citenamefont
  {Bertoldi}, \citenamefont {Landragin},\ and\ \citenamefont
  {Bouyer}}]{Bouyer11}%
  \BibitemOpen
  \bibfield  {author} {\bibinfo {author} {\bibfnamefont {S.}~\bibnamefont
  {Bernon}}, \bibinfo {author} {\bibfnamefont {T.}~\bibnamefont
  {Vanderbruggen}}, \bibinfo {author} {\bibfnamefont {R.}~\bibnamefont
  {Kohlhaas}}, \bibinfo {author} {\bibfnamefont {A.}~\bibnamefont {Bertoldi}},
  \bibinfo {author} {\bibfnamefont {A.}~\bibnamefont {Landragin}}, \ and\
  \bibinfo {author} {\bibfnamefont {P.}~\bibnamefont {Bouyer}},\ }\href
  {http://stacks.iop.org/1367-2630/13/i=6/a=065021} {\bibfield  {journal}
  {\bibinfo  {journal} {New Journal of Physics}\ }\textbf {\bibinfo {volume}
  {13}},\ \bibinfo {pages} {065021} (\bibinfo {year} {2011})}\BibitemShut
  {NoStop}%
\bibitem [{\citenamefont {M\"{o}ller}\ \emph {et~al.}(1998)\citenamefont
  {M\"{o}ller}, \citenamefont {Hoffer}, \citenamefont {Lippi}, \citenamefont
  {Ackemann}, \citenamefont {Gahl},\ and\ \citenamefont {Lange}}]{Lange98}%
  \BibitemOpen
  \bibfield  {author} {\bibinfo {author} {\bibfnamefont {M.}~\bibnamefont
  {M\"{o}ller}}, \bibinfo {author} {\bibfnamefont {L.~M.}\ \bibnamefont
  {Hoffer}}, \bibinfo {author} {\bibfnamefont {G.~L.}\ \bibnamefont {Lippi}},
  \bibinfo {author} {\bibfnamefont {T.}~\bibnamefont {Ackemann}}, \bibinfo
  {author} {\bibfnamefont {A.}~\bibnamefont {Gahl}}, \ and\ \bibinfo {author}
  {\bibfnamefont {W.}~\bibnamefont {Lange}},\ }\href {\doibase
  10.1080/09500349808231710} {\bibfield  {journal} {\bibinfo  {journal}
  {Journal of Modern Optics}\ }\textbf {\bibinfo {volume} {45}},\ \bibinfo
  {pages} {1913} (\bibinfo {year} {1998})}\BibitemShut {NoStop}%
\bibitem [{\citenamefont {Falke}\ \emph {et~al.}(2006)\citenamefont {Falke},
  \citenamefont {Tiemann}, \citenamefont {Lisdat}, \citenamefont {Schnatz},\
  and\ \citenamefont {Grosche}}]{Grosche06}%
  \BibitemOpen
  \bibfield  {author} {\bibinfo {author} {\bibfnamefont {S.}~\bibnamefont
  {Falke}}, \bibinfo {author} {\bibfnamefont {E.}~\bibnamefont {Tiemann}},
  \bibinfo {author} {\bibfnamefont {C.}~\bibnamefont {Lisdat}}, \bibinfo
  {author} {\bibfnamefont {H.}~\bibnamefont {Schnatz}}, \ and\ \bibinfo
  {author} {\bibfnamefont {G.}~\bibnamefont {Grosche}},\ }\href {\doibase
  10.1103/PhysRevA.74.032503} {\bibfield  {journal} {\bibinfo  {journal} {Phys.
  Rev. A}\ }\textbf {\bibinfo {volume} {74}},\ \bibinfo {pages} {032503}
  (\bibinfo {year} {2006})}\BibitemShut {NoStop}%
\bibitem [{\citenamefont {Wang}\ \emph {et~al.}(1997)\citenamefont {Wang},
  \citenamefont {Gould},\ and\ \citenamefont {Stwalley}}]{Stwalley97}%
  \BibitemOpen
  \bibfield  {author} {\bibinfo {author} {\bibfnamefont {H.}~\bibnamefont
  {Wang}}, \bibinfo {author} {\bibfnamefont {P.~L.}\ \bibnamefont {Gould}}, \
  and\ \bibinfo {author} {\bibfnamefont {W.~C.}\ \bibnamefont {Stwalley}},\
  }\href {\doibase 10.1063/1.473804} {\bibfield  {journal} {\bibinfo  {journal}
  {The Journal of Chemical Physics}\ }\textbf {\bibinfo {volume} {106}},\
  \bibinfo {pages} {7899} (\bibinfo {year} {1997})}\BibitemShut {NoStop}%
\bibitem [{\citenamefont {Carmichael}\ and\ \citenamefont
  {Sanders}(1999)}]{Carmichael99}%
  \BibitemOpen
  \bibfield  {author} {\bibinfo {author} {\bibfnamefont {H.~J.}\ \bibnamefont
  {Carmichael}}\ and\ \bibinfo {author} {\bibfnamefont {B.~C.}\ \bibnamefont
  {Sanders}},\ }\href {\doibase 10.1103/PhysRevA.60.2497} {\bibfield  {journal}
  {\bibinfo  {journal} {Phys. Rev. A}\ }\textbf {\bibinfo {volume} {60}},\
  \bibinfo {pages} {2497} (\bibinfo {year} {1999})}\BibitemShut {NoStop}%
\bibitem [{\citenamefont {Gripp}\ \emph {et~al.}(1997)\citenamefont {Gripp},
  \citenamefont {Mielke},\ and\ \citenamefont {Orozco}}]{Orozco97}%
  \BibitemOpen
  \bibfield  {author} {\bibinfo {author} {\bibfnamefont {J.}~\bibnamefont
  {Gripp}}, \bibinfo {author} {\bibfnamefont {S.~L.}\ \bibnamefont {Mielke}}, \
  and\ \bibinfo {author} {\bibfnamefont {L.~A.}\ \bibnamefont {Orozco}},\
  }\href {\doibase 10.1103/PhysRevA.56.3262} {\bibfield  {journal} {\bibinfo
  {journal} {Phys. Rev. A}\ }\textbf {\bibinfo {volume} {56}},\ \bibinfo
  {pages} {3262} (\bibinfo {year} {1997})}\BibitemShut {NoStop}%
\bibitem [{\citenamefont {Scully}(1991)}]{Scully91}%
  \BibitemOpen
  \bibfield  {author} {\bibinfo {author} {\bibfnamefont {M.~O.}\ \bibnamefont
  {Scully}},\ }\href {\doibase 10.1103/PhysRevLett.67.1855} {\bibfield
  {journal} {\bibinfo  {journal} {Phys. Rev. Lett.}\ }\textbf {\bibinfo
  {volume} {67}},\ \bibinfo {pages} {1855} (\bibinfo {year}
  {1991})}\BibitemShut {NoStop}%
\bibitem [{\citenamefont {Lampis}\ \emph {et~al.}(2016)\citenamefont {Lampis},
  \citenamefont {Culver}, \citenamefont {Megyeri},\ and\ \citenamefont
  {Goldwin}}]{Lampis16}%
  \BibitemOpen
  \bibfield  {author} {\bibinfo {author} {\bibfnamefont {A.}~\bibnamefont
  {Lampis}}, \bibinfo {author} {\bibfnamefont {R.}~\bibnamefont {Culver}},
  \bibinfo {author} {\bibfnamefont {B.}~\bibnamefont {Megyeri}}, \ and\
  \bibinfo {author} {\bibfnamefont {J.}~\bibnamefont {Goldwin}},\ }\href
  {\doibase 10.1364/OE.24.015494} {\bibfield  {journal} {\bibinfo  {journal}
  {Opt. Express}\ }\textbf {\bibinfo {volume} {24}},\ \bibinfo {pages} {15494}
  (\bibinfo {year} {2016})}\BibitemShut {NoStop}%
\bibitem [{\citenamefont {Gokhroo}\ \emph {et~al.}(2011)\citenamefont
  {Gokhroo}, \citenamefont {Rajalakshmi}, \citenamefont {Easwaran},\ and\
  \citenamefont {Unnikrishnan}}]{Mumbai11}%
  \BibitemOpen
  \bibfield  {author} {\bibinfo {author} {\bibfnamefont {V.}~\bibnamefont
  {Gokhroo}}, \bibinfo {author} {\bibfnamefont {G.}~\bibnamefont
  {Rajalakshmi}}, \bibinfo {author} {\bibfnamefont {R.~K.}\ \bibnamefont
  {Easwaran}}, \ and\ \bibinfo {author} {\bibfnamefont {C.~S.}\ \bibnamefont
  {Unnikrishnan}},\ }\href {http://stacks.iop.org/0953-4075/44/i=11/a=115307}
  {\bibfield  {journal} {\bibinfo  {journal} {Journal of Physics B: Atomic,
  Molecular and Optical Physics}\ }\textbf {\bibinfo {volume} {44}},\ \bibinfo
  {pages} {115307} (\bibinfo {year} {2011})}\BibitemShut {NoStop}%
\bibitem [{\citenamefont {Landini}\ \emph {et~al.}(2011)\citenamefont
  {Landini}, \citenamefont {Roy}, \citenamefont {Carcagn\'{\i}}, \citenamefont
  {Trypogeorgos}, \citenamefont {Fattori}, \citenamefont {Inguscio},\ and\
  \citenamefont {Modugno}}]{Modugno11}%
  \BibitemOpen
  \bibfield  {author} {\bibinfo {author} {\bibfnamefont {M.}~\bibnamefont
  {Landini}}, \bibinfo {author} {\bibfnamefont {S.}~\bibnamefont {Roy}},
  \bibinfo {author} {\bibfnamefont {L.}~\bibnamefont {Carcagn\'{\i}}}, \bibinfo
  {author} {\bibfnamefont {D.}~\bibnamefont {Trypogeorgos}}, \bibinfo {author}
  {\bibfnamefont {M.}~\bibnamefont {Fattori}}, \bibinfo {author} {\bibfnamefont
  {M.}~\bibnamefont {Inguscio}}, \ and\ \bibinfo {author} {\bibfnamefont
  {G.}~\bibnamefont {Modugno}},\ }\href {\doibase 10.1103/PhysRevA.84.043432}
  {\bibfield  {journal} {\bibinfo  {journal} {Phys. Rev. A}\ }\textbf {\bibinfo
  {volume} {84}},\ \bibinfo {pages} {043432} (\bibinfo {year}
  {2011})}\BibitemShut {NoStop}%
\bibitem [{\citenamefont {Marangoni}\ \emph {et~al.}(2012)\citenamefont
  {Marangoni}, \citenamefont {Menegatti},\ and\ \citenamefont
  {Marcassa}}]{Marcassa12}%
  \BibitemOpen
  \bibfield  {author} {\bibinfo {author} {\bibfnamefont {B.~S.}\ \bibnamefont
  {Marangoni}}, \bibinfo {author} {\bibfnamefont {C.~R.}\ \bibnamefont
  {Menegatti}}, \ and\ \bibinfo {author} {\bibfnamefont {L.~G.}\ \bibnamefont
  {Marcassa}},\ }\href {http://stacks.iop.org/0953-4075/45/i=17/a=175301}
  {\bibfield  {journal} {\bibinfo  {journal} {Journal of Physics B: Atomic,
  Molecular and Optical Physics}\ }\textbf {\bibinfo {volume} {45}},\ \bibinfo
  {pages} {175301} (\bibinfo {year} {2012})}\BibitemShut {NoStop}%
\bibitem [{\citenamefont {Laupr\^{e}tre}\ \emph {et~al.}(2011)\citenamefont
  {Laupr\^{e}tre}, \citenamefont {Proux}, \citenamefont {Ghosh}, \citenamefont
  {Schwartz}, \citenamefont {Goldfarb},\ and\ \citenamefont
  {Bretenaker}}]{Laupretre11}%
  \BibitemOpen
  \bibfield  {author} {\bibinfo {author} {\bibfnamefont {T.}~\bibnamefont
  {Laupr\^{e}tre}}, \bibinfo {author} {\bibfnamefont {C.}~\bibnamefont
  {Proux}}, \bibinfo {author} {\bibfnamefont {R.}~\bibnamefont {Ghosh}},
  \bibinfo {author} {\bibfnamefont {S.}~\bibnamefont {Schwartz}}, \bibinfo
  {author} {\bibfnamefont {F.}~\bibnamefont {Goldfarb}}, \ and\ \bibinfo
  {author} {\bibfnamefont {F.}~\bibnamefont {Bretenaker}},\ }\href {\doibase
  10.1364/OL.36.001551} {\bibfield  {journal} {\bibinfo  {journal} {Opt.
  Lett.}\ }\textbf {\bibinfo {volume} {36}},\ \bibinfo {pages} {1551} (\bibinfo
  {year} {2011})}\BibitemShut {NoStop}%
\bibitem [{\citenamefont {Hilico}\ \emph {et~al.}(1992)\citenamefont {Hilico},
  \citenamefont {Fabre},\ and\ \citenamefont {Giacobino}}]{Giacobino92}%
  \BibitemOpen
  \bibfield  {author} {\bibinfo {author} {\bibfnamefont {L.}~\bibnamefont
  {Hilico}}, \bibinfo {author} {\bibfnamefont {C.}~\bibnamefont {Fabre}}, \
  and\ \bibinfo {author} {\bibfnamefont {E.}~\bibnamefont {Giacobino}},\ }\href
  {http://stacks.iop.org/0295-5075/18/i=8/a=004} {\bibfield  {journal}
  {\bibinfo  {journal} {EPL (Europhysics Letters)}\ }\textbf {\bibinfo {volume}
  {18}},\ \bibinfo {pages} {685} (\bibinfo {year} {1992})}\BibitemShut
  {NoStop}%
\bibitem [{\citenamefont {Guerin}\ \emph {et~al.}(2008)\citenamefont {Guerin},
  \citenamefont {Michaud},\ and\ \citenamefont {Kaiser}}]{Kaiser08}%
  \BibitemOpen
  \bibfield  {author} {\bibinfo {author} {\bibfnamefont {W.}~\bibnamefont
  {Guerin}}, \bibinfo {author} {\bibfnamefont {F.}~\bibnamefont {Michaud}}, \
  and\ \bibinfo {author} {\bibfnamefont {R.}~\bibnamefont {Kaiser}},\ }\href
  {\doibase 10.1103/PhysRevLett.101.093002} {\bibfield  {journal} {\bibinfo
  {journal} {Phys. Rev. Lett.}\ }\textbf {\bibinfo {volume} {101}},\ \bibinfo
  {pages} {093002} (\bibinfo {year} {2008})}\BibitemShut {NoStop}%
\bibitem [{\citenamefont {Vrijsen}\ \emph {et~al.}(2011)\citenamefont
  {Vrijsen}, \citenamefont {Hosten}, \citenamefont {Lee}, \citenamefont
  {Bernon},\ and\ \citenamefont {Kasevich}}]{Kasevich11}%
  \BibitemOpen
  \bibfield  {author} {\bibinfo {author} {\bibfnamefont {G.}~\bibnamefont
  {Vrijsen}}, \bibinfo {author} {\bibfnamefont {O.}~\bibnamefont {Hosten}},
  \bibinfo {author} {\bibfnamefont {J.}~\bibnamefont {Lee}}, \bibinfo {author}
  {\bibfnamefont {S.}~\bibnamefont {Bernon}}, \ and\ \bibinfo {author}
  {\bibfnamefont {M.~A.}\ \bibnamefont {Kasevich}},\ }\href {\doibase
  10.1103/PhysRevLett.107.063904} {\bibfield  {journal} {\bibinfo  {journal}
  {Phys. Rev. Lett.}\ }\textbf {\bibinfo {volume} {107}},\ \bibinfo {pages}
  {063904} (\bibinfo {year} {2011})}\BibitemShut {NoStop}%
\bibitem [{\citenamefont {Bohnet}\ \emph {et~al.}(2012)\citenamefont {Bohnet},
  \citenamefont {Chen}, \citenamefont {Weiner}, \citenamefont {Meiser},
  \citenamefont {Holland},\ and\ \citenamefont {Thompson}}]{Thompson12}%
  \BibitemOpen
  \bibfield  {author} {\bibinfo {author} {\bibfnamefont {J.~G.}\ \bibnamefont
  {Bohnet}}, \bibinfo {author} {\bibfnamefont {Z.}~\bibnamefont {Chen}},
  \bibinfo {author} {\bibfnamefont {J.~M.}\ \bibnamefont {Weiner}}, \bibinfo
  {author} {\bibfnamefont {D.}~\bibnamefont {Meiser}}, \bibinfo {author}
  {\bibfnamefont {M.~J.}\ \bibnamefont {Holland}}, \ and\ \bibinfo {author}
  {\bibfnamefont {J.~K.}\ \bibnamefont {Thompson}},\ }\href {\doibase
  10.1038/nature10920} {\bibfield  {journal} {\bibinfo  {journal} {Nature}\
  }\textbf {\bibinfo {volume} {484}},\ \bibinfo {pages} {78} (\bibinfo {year}
  {2012})}\BibitemShut {NoStop}%
\bibitem [{\citenamefont {Norcia}\ and\ \citenamefont
  {Thompson}(2016)}]{Thompson16}%
  \BibitemOpen
  \bibfield  {author} {\bibinfo {author} {\bibfnamefont {M.~A.}\ \bibnamefont
  {Norcia}}\ and\ \bibinfo {author} {\bibfnamefont {J.~K.}\ \bibnamefont
  {Thompson}},\ }\href {\doibase 10.1103/PhysRevX.6.011025} {\bibfield
  {journal} {\bibinfo  {journal} {Phys. Rev. X}\ }\textbf {\bibinfo {volume}
  {6}},\ \bibinfo {pages} {011025} (\bibinfo {year} {2016})}\BibitemShut
  {NoStop}%
\bibitem [{\citenamefont {Kuppens}\ \emph {et~al.}(1994)\citenamefont
  {Kuppens}, \citenamefont {van Exter},\ and\ \citenamefont
  {Woerdman}}]{Woerdman94}%
  \BibitemOpen
  \bibfield  {author} {\bibinfo {author} {\bibfnamefont {S.~J.~M.}\
  \bibnamefont {Kuppens}}, \bibinfo {author} {\bibfnamefont {M.~P.}\
  \bibnamefont {van Exter}}, \ and\ \bibinfo {author} {\bibfnamefont {J.~P.}\
  \bibnamefont {Woerdman}},\ }\href {\doibase 10.1103/PhysRevLett.72.3815}
  {\bibfield  {journal} {\bibinfo  {journal} {Phys. Rev. Lett.}\ }\textbf
  {\bibinfo {volume} {72}},\ \bibinfo {pages} {3815} (\bibinfo {year}
  {1994})}\BibitemShut {NoStop}%
\bibitem [{\citenamefont {Laupr\^{e}tre}\ \emph {et~al.}(2012)\citenamefont
  {Laupr\^{e}tre}, \citenamefont {Schwartz}, \citenamefont {Ghosh},
  \citenamefont {Carusotto}, \citenamefont {Goldfarb},\ and\ \citenamefont
  {Bretenaker}}]{Laupretre12}%
  \BibitemOpen
  \bibfield  {author} {\bibinfo {author} {\bibfnamefont {T.}~\bibnamefont
  {Laupr\^{e}tre}}, \bibinfo {author} {\bibfnamefont {S.}~\bibnamefont
  {Schwartz}}, \bibinfo {author} {\bibfnamefont {R.}~\bibnamefont {Ghosh}},
  \bibinfo {author} {\bibfnamefont {I.}~\bibnamefont {Carusotto}}, \bibinfo
  {author} {\bibfnamefont {F.}~\bibnamefont {Goldfarb}}, \ and\ \bibinfo
  {author} {\bibfnamefont {F.}~\bibnamefont {Bretenaker}},\ }\href
  {http://stacks.iop.org/1367-2630/14/i=4/a=043012} {\bibfield  {journal}
  {\bibinfo  {journal} {New Journal of Physics}\ }\textbf {\bibinfo {volume}
  {14}},\ \bibinfo {pages} {043012} (\bibinfo {year} {2012})}\BibitemShut
  {NoStop}%
\bibitem [{\citenamefont {Shahriar}\ \emph {et~al.}(2007)\citenamefont
  {Shahriar}, \citenamefont {Pati}, \citenamefont {Tripathi}, \citenamefont
  {Gopal}, \citenamefont {Messall},\ and\ \citenamefont {Salit}}]{Shahriar07}%
  \BibitemOpen
  \bibfield  {author} {\bibinfo {author} {\bibfnamefont {M.~S.}\ \bibnamefont
  {Shahriar}}, \bibinfo {author} {\bibfnamefont {G.~S.}\ \bibnamefont {Pati}},
  \bibinfo {author} {\bibfnamefont {R.}~\bibnamefont {Tripathi}}, \bibinfo
  {author} {\bibfnamefont {V.}~\bibnamefont {Gopal}}, \bibinfo {author}
  {\bibfnamefont {M.}~\bibnamefont {Messall}}, \ and\ \bibinfo {author}
  {\bibfnamefont {K.}~\bibnamefont {Salit}},\ }\href {\doibase
  10.1103/PhysRevA.75.053807} {\bibfield  {journal} {\bibinfo  {journal} {Phys.
  Rev. A}\ }\textbf {\bibinfo {volume} {75}},\ \bibinfo {pages} {053807}
  (\bibinfo {year} {2007})}\BibitemShut {NoStop}%
\bibitem [{dat()}]{data}%
  \BibitemOpen
  \href {http://epapers.bham.ac.uk/2212/} {}\bibinfo {note}
  {\lowercase{h}ttp://epapers.bham.ac.uk/2212/}\BibitemShut {NoStop}%
\end{thebibliography}%

\end{document}